\documentclass[journal,final,twocolumn]{elsarticle}

\usepackage{lineno,hyperref}
\usepackage{tabularx}
\usepackage{paralist}
\usepackage{subfigure}
\usepackage{mdwlist}
\usepackage{color}
\usepackage{amsmath}

\modulolinenumbers[5]

\journal{Sensors and Actuators A: Physical}

\begin{document}
\newcolumntype{L}[1]{>{\raggedright\arraybackslash}p{#1}}
\newcolumntype{C}[1]{>{\centering\arraybackslash}p{#1}}
\newcolumntype{R}[1]{>{\raggedleft\arraybackslash}p{#1}}

\begin{frontmatter}

\title{Piezoresistive Foam Sensor Arrays for Marine Applications}

\author[mymainaddress]{Jeff E. Dusek\corref{mycorrespondingauthor}\fnref{fn1}}
\cortext[mycorrespondingauthor]{Corresponding author}
\ead{jdusek1@mit.edu}
\fntext[fn1]{This work is based on a portion of the author's doctoral thesis \cite{Dusek:2015p1}.}

\author[mymainaddress]{Michael S. Triantafyllou}

\author[mysecondaryaddress]{Jeffrey H. Lang}

\address[mymainaddress]{Department of Mechanical Engineering, Massachusetts Institute of Technology, 77 Massachusetts Ave., Cambridge, Massachusetts 02139}
\address[mysecondaryaddress]{Electrical Engineering and Computer Science Department, Massachusetts Institute of Technology, 77 Massachusetts Ave., Cambridge, Massachusetts 02139}

\begin{abstract}

Spatially-dense pressure measurements are needed on curved surfaces in marine environments to provide marine vehicles with the detailed, real-time measurements of the near-field flow necessary to improve performance through flow control. To address this challenge, a waterproof and conformal pressure sensor array comprising carbon black-doped-silicone closed-cell foam (CBPDMS foam) was developed for use in marine applications. The response of the CBPDMS foam sensor arrays was characterized using periodic hydrodynamic pressure stimuli from vertical plunging, from which a piecewise polynomial calibration was developed to describe the sensor response. Inspired by the distributed pressure and velocity sensing capabilities of the fish lateral line, the CBPDMS foam sensor arrays have significant advantages over existing commercial sensors for distributed flow reconstruction and control. Experimental results have shown the sensor arrays to have sensitivity on the order of 5 Pascal, dynamic range of 50-500 Pascal; are contained in a waterproof and completely flexible package, and have material cost less than $\$$10 per sensor. 

\end{abstract}


\end{frontmatter}

%\linenumbers
\let\thefootnote\relax\footnotetext{\textbf{Abbreviations} CB: Carbon Black, PDMS: Polydimethylsiloxane, CBPDMS: Carbon black-PDMS Composite, Ag-CBPDMS: Silver-carbon black-PDMS composite}
\addtocounter{footnote}{-1}\let\thefootnote\svthefootnote

\section{Introduction}

Operating in the marine environment places unique challenges on unmanned systems that require novel approaches to sensing and control. Marine vehicle performance is largely influenced by interactions with the flow around their hull, both self-generated and environmentally-driven. To improve performance through flow control, a detailed, real-time measurement of the near-field flow is necessary, yet such sensing capability is presently unavailable. In nature, fish have overcome this sensory deficit by utilizing feedback from the lateral line sensory organ. Comprising two subsystems that act as distributed velocity and pressure sensor arrays, respectively, the lateral line has been shown to mediate such complex behaviors as obstacle detection and avoidance, organism tracking, vortex wake synchronization, and cooperative schooling. When considering the ability of fish to thrive in the inhospitable marine environment, achieving even a portion of their capabilities with engineering systems would have tremendous benefits to oceanographic research, marine vehicle performance, and undersea exploration. To address the challenges associated with obtaining spatially-dense pressure measurements on curved surfaces in marine environments, a new waterproof and conformal pressure sensor array was developed based on a closed-cell piezoresistive foam comprising carbon black-doped-silicone composite (CBPDMS foam). 

\subsection{A Biological Solution: The Lateral Line}

Solutions to complex engineering problems can often be found by looking to nature for inspiration, especially when considering the marine environment which is inhospitable to humans, but contains an incredible degree of bio-diversity. The ability of fish to navigate the undersea world at high speeds and in close proximity to obstacles and other individuals is particularly attractive to ocean engineers seeking to enhance the performance of marine vehicles. The lateral line system found in all species of fish and some amphibians is a hair-cell based mechanosensory organ comprising two primary subsystems: superficial neuromasts and canal neuromasts. While the two subsystems share physiological elements, they provide different information to the fish about the surrounding fluid flow, allowing for a wide variety of behaviors.

\subsubsection{Lateral Line Physiology}

The main building block of the distributed lateral line system is neuromasts, small epithelial organs which comprises both mechanosensory hair cells and non-sensory cells. Superficial neuromasts are located on the surface of the skin and are generally considered as velocity sensors \cite{Bleckmann:2012p4730,Coombs:2014p4945}, where stimulation is caused by deflections due to viscous drag \cite{Windsor:2009p4593}.

The canal subsystem comprises a series of canals beneath the skin and connected to the surface through an array of pores.  When exposed to a pressure gradient between adjacent pores, a local flow is induced in the canal, stimulating canal neuromasts located within the canal and between the pores \cite{Windsor:2009p4593,Rapo:2009p442}. The deflection of the canal neuromasts is proportional to the velocity within the canal, which is proportional to the pressure difference between pores \cite{Windsor:2009p4593,Rapo:2009p442}.  In this way, the trunk canal subsystem acts in a similar fashion to an array of differential pressure sensors mounted on the animal's body in hydrodynamically sensitive regions \cite{Fernandez:2011p4634}. 

\section{Bringing Lateral Line Inspired Capabilities to Marine Vehicles}
\label{sec:vehicles}

In order to achieve the lateral line mediated capabilities observed in fish with marine vehicles, advancements are needed in the characterization and utilization of near-body hydrodynamic pressure signals, as well as in the development of bio-inspired distributed sensor arrays. The fish lateral line and the behaviors it mediates provides an excellent motivation for the development of distributed sensor arrays, however the design requirements for arrays intended for marine vehicle use must be scaled appropriately to the hydrodynamic stimuli of interest.

To elucidate the design guidelines for a distributed hydrodynamic pressure sensor array, a series of towing tank and field experiments were conducted using arrays of commercially available pressure sensors. Experiments with an instrumented model sailboat hull of length L=1 m were conducted in the MIT towing tank \cite{Fernandez:2011p4407}. An array of 10 Honeywell 19mm series pressure sensors were connected to taps along the length of the model's hull and it was found that the pressure measurements could be used to accurately estimate the hull's angle of attack. Similarly, Honeywell 19mm pressure sensors were utilized to study the formation and shedding of leading edge vortices from a hydrofoil towed at large angles of attack \cite{Dusek:2011p4406}. The utility of near-body flow sensing on a surface vehicle was investigated experimentally using an unmanned kayak vehicle equipped with an array of 20 Honeywell SPT series pressure sensors mounted inside the vehicle hull, necessitating the drilling of holes through the hull to access the flow \cite{Dusek:2013p4635}. It was found that the vehicle dynamics in pitch and roll were measurable by the pressure sensor array, as well as the initiation of vehicle maneuvers due to a characteristic pressure signature consistent with added mass effects during unsteady vehicle motions. The magnitude and frequency of dynamic pressure signals measured during these experiments, and analysis of the sensor spacing on the foil and hull surfaces, lead to the development of design guidelines for distributed pressure sensor arrays given in Section \ref{sec:guidelines}. 

\subsection{Guidelines for Hydrodynamic Pressure Sensor Arrays}
\label{sec:guidelines}

In general, the sensors used in the studies outlined in Section \ref{sec:vehicles} were rigid and too large for surface mounting applications, and were not designed for prolonged exposure to moisture. Based on the results of these studies, the attributes of a pressure sensor array designed specifically for marine use were developed as follows. 

\begin{compactdesc}
\item[Flexibility] Pressure sensor arrays to be surface mounted on curved surfaces with radius of curvature $\sim$5-10 cm, consistent with unmanned vehicle hulls \cite{Dusek:2011p4406,Dusek:2013p4635,Fernandez:2011p4407}.
\item[Form Factor] Sensor spacing should be less than 5 cm for vehicles on the 1-3 m scale. Sensor thickness should be $<$ 5 mm to avoid vortex shedding \cite{Dusek:2011p4406,Fernandez:2011p4634,Fernandez:2011p4407}.
\item[Robustness]  Sensor arrays are meant for sustained underwater or exposed operation and may be subject to impacts.
\item[Dynamic Range] For marine vehicles of length $\sim$1-3 m, dynamic pressure stimuli range from $\sim$10-400 Pa \cite{Dusek:2011p4406,Dusek:2013p4635,Fernandez:2011p4634,Fernandez:2011p4407}.
\item[Sensitivity] For marine vehicles of length $\sim$1-3 m, sensitivity of $\sim$10 Pa is desired to characterize hydrodynamic stimuli \cite{Dusek:2011p4406,Dusek:2013p4635,Fernandez:2011p4634,Fernandez:2011p4407}.
\item[Cost] Sensor cost should be reduced from $\sim\$$100 per sensor to $\sim\$$10 per sensor \cite{Dusek:2013p4635}.
\end{compactdesc}

\subsection{Doped-Polymer `Smart-Skins' for Marine Applications}
Based on the operational requirements for use in distributed pressure sensing on marine vehicles, doped polymer `smart-skin' arrays offer the best combination of performance characteristics for marine use. Doped polymers allow for the development of a completely waterproof, stretchable, and flexible sensor array through the careful selection of bulk matrix material and conductive dopant. The flexibility, robustness, and resistance to moisture of bulk matrix materials like PDMS (silicone) make doped composites well suited for prolonged environmental exposure while surface mounted on marine vehicle hulls. Additionally, controlling the material properties of the bulk matrix through the introduction of porosity has been shown to increase sensitivity in carbon black-PDMS composites, allowing for piezoresistive composites to be optimized for pressure ranges consistent with hydrodynamic stimulus \cite{Dusek:2014p4920}. Finally, doped polymers make use of cheap and easy to work with component materials, allowing for the scaling of distributed pressure sensor arrays to spatially-dense applications.

\section{Carbon Black-PDMS Composite Sensor Development}
\label{Sec:Sensor_Development}

Carbon black-PDMS foam pressure sensor arrays rely on the piezoresitivity, or variation in resistance with strain, of carbon black doped PDMS (silicone) to provide an indirect measurement of pressure stimulus. Carbon black-PDMS (CBPDMS) composite has been studied as an active material for pressure and shear sensors due to its low Young's modulus, ease of fabrication, and low cost \cite{Luheng:2009p4656, Lotters:1997p147, Yaul:2012p4698, Lacasse:2010p4697, Ryvkina:2005p4655}. Each component composite in the sensor array utilizes PDMS as the matrix material, ensuring strong bonding between sensor components while retaining overall array flexibility. By varying the carbon black doping and Young's modulus between sections of the sensor array, a linear four by one array of sensor channels was created here. Electrodes were fabricated using a silver-carbon black-PDMS (Ag-CBPDMS) composite with a high mass fraction of silver to ensure high conductance and low piezoresistivity, as a discussed in Section \ref{Sec:electrode}. Sensing elements were fabricated using a closed-cell CBPDMS foam (Section \ref{Sec:foam}) with a low Young's modulus and carbon black concentration near the percolation threshold to enhance piezoresistivity and improve the sensitivity of the array to hydrodynamic stimuli.

\subsection{Models of CBPDMS Piezoresistivity}
\label{Sec:models}

Above a mass fraction of carbon black known as the percolation threshold, carbon black doped PDMS forms a conductive composite. When the percolation threshold is reached, continuous chains of carbon black particles create conductive pathways through the PDMS bulk material; while below the percolation threshold, CBPDMS composite behaves like an insulator. The piezoresistive behavior of CBPDMS relies on the formation and breakdown of these conductive chains as the PDMS matrix material is deformed due to external stimulus. 

The resistance change of CBPDMS composite when subjected to an external pressure stimulus has been primarily described using two models, the compressible and incompressible models \cite{Lacasse:2010p4697}. Both piezoresistive effects have been observed in CBPDMS composites, and the relationship between resistance and strain is highly dependent on the filler material, the type of polymer matrix, and the nature of the loading \cite{Lacasse:2010p4697}. For the CBPDMS foam pressure sensor array using four-point probe measurements and responding to isotropic hydrodynamic stimuli, the resistance change was found to be consistent with the compressible model.

\begin{figure}
        \centering
\includegraphics[width=6cm]{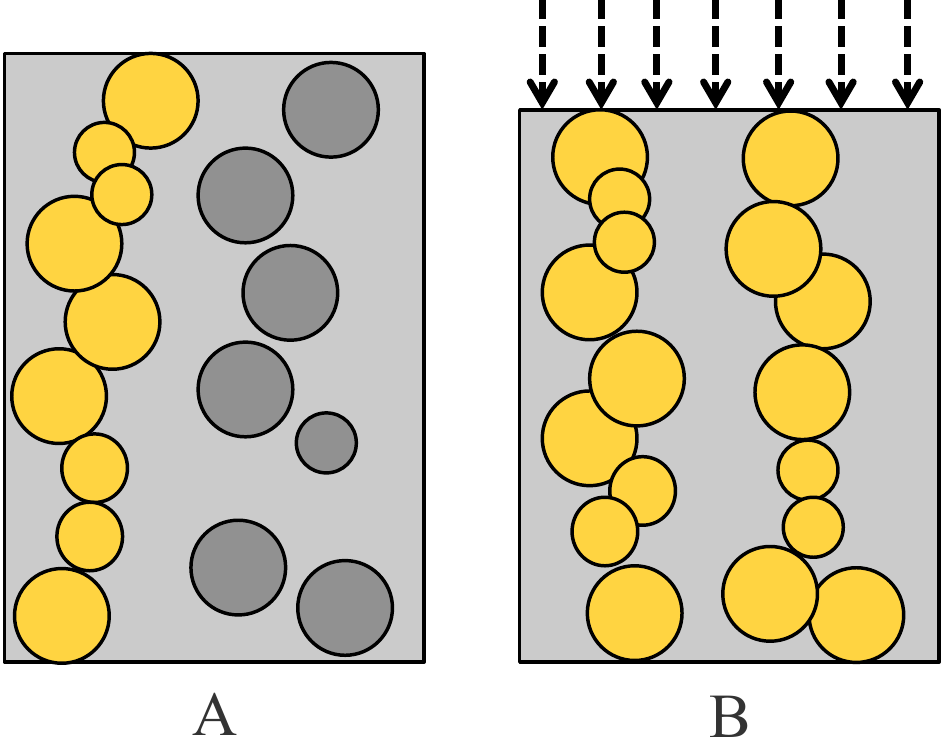}
 \caption[The Compressible Model]{The compressible model of CBPDMS piezoresistivity in which the PDMS matrix is compressed under load, leading to an increased volume fraction of CB particles in the composite and the formation of new conductive pathways, decreasing resistance. The light grey boxes represent the PDMS foam bulk material, and dark grey circles represent free carbon black (CB) particles that have not formed conductive chains. The gold circles represent CB particles that have linked together to form continuous conductive pathways through the bulk PDMS foam matrix.}
 \label{fig:compressible_model}
\end{figure}

The compressible model of CBPMS piezoresistivity, shown in Figure \ref{fig:compressible_model}, states that as the PDMS matrix is compressed due to external pressure stimulus, the volume fraction of CB particles within the composite is increased, allowing for the formation of new continuous conductive pathways and decreasing the resistance of the composite \cite{Lacasse:2010p4697, Hussain:2001p1689, Beruto:2005p301}. When the external pressure stimulus on the composite is decreased, the material relaxes, and the newly formed conductive pathways are broken, leading to a recovery of the composite resistance.

\subsection{DC Response of CBPDMS Composite}

Carbon black-PDMS composites provide a reversible piezoresistive response to dynamic pressure stimulus, but are not well suited for low frequency, DC type stimulus. Over time at a continuous strain, conductive pathways relax and resistance changes, effectively acting as a high pass filter \cite{Ding:2007p4912}. Efforts have been made to characterize the low frequency behavior of CBDPMS composites through the use of Burger models and modifications of the Burger model by Yaul  \cite{Yaul:2012p4698} and Lacasse \cite{Lacasse:2010p4697}. The hydrodynamic pressure stimuli of interest to marine vehicles typically have frequencies ranging from 0.5 Hz (water waves, vortex shedding) to 35 Hz (predator and prey detection), and the dynamic response of the CBPDMS foam sensor arrays will be the focus of this study.

\subsection{Addition of Porosity to CBPDMS for Increased Sensitivity}
\label{sec:Porous CBPDMS}

Version one `smart-skin' pressure sensors were developed utilizing a solid CBPDMS composite arranged in a strain-gauge arrangement on solid PDMS substrates \cite{Yaul:2012p4698, Fernandez:2011p3057}. To achieve the sensitivity required for the measurement of hydrodynamic stimuli, strain enhancing diaphragms were molded into the PDMS substrates, complicating the sensor fabrication and reducing array reliability.  To achieve the sensitivity provided by the strain enhancing diaphragms without the need for additional molded features, the introduction of porosity into PDMS through the use of a sacrificial sugar scaffold, as demonstrated by King and Cha \cite{King:2009p4657, Cha:2011p4658}, was chosen as a viable method to lower Young's modulus and increase piezoresistivity.

Using a sacrificial sugar scaffold with CBPDMS composite, version two open-cell CBPDMS foam pressure sensor arrays were developed for hydrodynamic sensing applications \cite{Dusek:2014p4920}. The Young's modulus of the CBPDMS active material was reduced by two orders of magnitude, leading to a 25 times increase in sensitivity over the version one strain-gauge based `smart-skin' arrays. The open-cell CBPDS foam sensor arrays were tested in the MIT Towing Tank using the dynamic pressure signal generated by surface water waves, demonstrating the utility of the arrays in hydrodynamic sensing applications.

The benefits of the strain enhancing diaphragms were largely retained in a robust and flexible simplified geometry by introducing porosity into the CBPDMS composite to create an open-cell foam \cite{Dusek:2014p4920}.However, the introduction of porosity introduced different problems, particularly the time consuming need to soak the CBPDMS for 24 hours to dissolve the sugar scaffold, and the requirement to waterproof the open-cell foam. It was also found that the version two CBPDMS open cell foam sensors were thicker than desired, and required special attention to be paid to electrical connections because of the reduced contact area between the porous material and the wires. Despite these challenges, the improvement in sensitivity shown by the CBPDMS foam motivated a continued effort to develop the material for use in hydrodynamic pressure sensing applications.

\subsection{Sensor Array Development Goals}

The goals of version three sensor development are summarized below:
\begin{compactenum} 
\item{Retain or improve upon sensitivity of open-cell CBPDMS foam}
\item{Eliminate the need for additional waterproofing layers}
\item{Improve contact between electrodes and piezoresistive material}
\item{Improve robustness of the sensor array and associated electrical wiring}
\end{compactenum}

\section{Closed-Cell CBDPMS Foam Sensor Arrays}

\begin{figure}
        \centering
        \subfigure[The top of the CBPDMS foam sensor array was covered with a thin layer of CBPDMS foam. The ends of a ribbon cable were embedded in a Ag-CBPDMS electrodes during fabrication.]{
                \includegraphics[width=8cm]{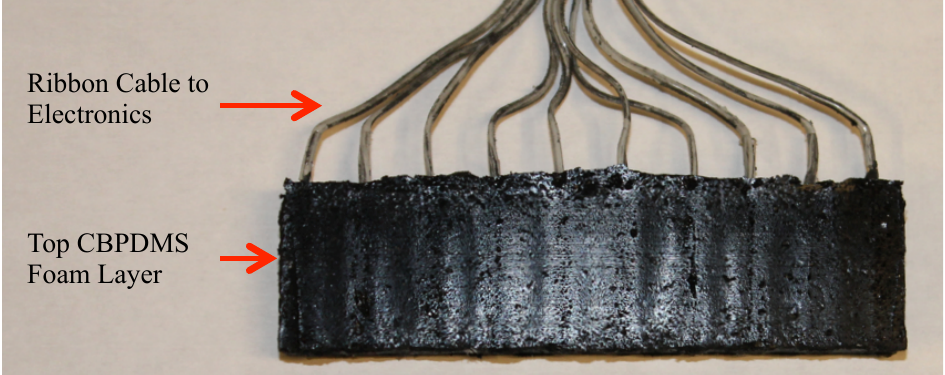}
                \label{fig:sensor_top}
                }
                  \subfigure[The overall footprint of the CBPDMS foam array was 80 mm by 20mm, and each sensor channel was 5 mm wide while electrodes and non-sensor foam sections were 4 mm wide]{
                \includegraphics[width=8cm]{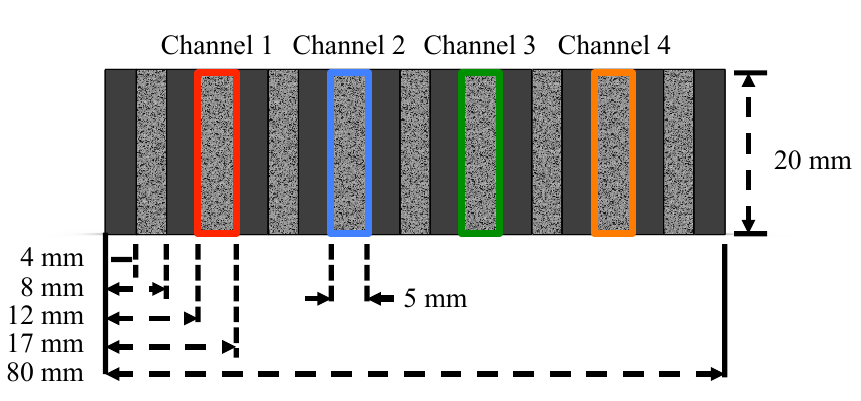}
                \label{fig:dimensions_top}
     		 }
                     \subfigure[Each CBPDMS foam array was composed of three layers. A 2 mm layer of pure PDMS provided a flexible base to the array. The active middle layer was composed of CBPDMS foam sensor channels and Ag-CBPDMS electrodes.The top layer of the device was a 1 mm thick layer of CBPDMS foam created during the foam expansion process.]{
                \includegraphics[width=8cm]{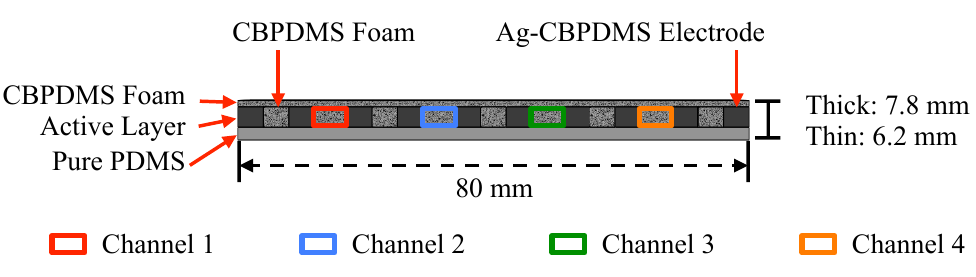}
                \label{fig:dimensions_front}
                }
                \caption[Closed-cell CBPDMS foam sensor arrays]{Carbon black- PDMS closed-cell foam pressure sensor arrays were fabricated using alternating segments of piezoresistive CBPDMS foam and silver-carbon black-PDMS (Ag-CBPDMS) electrodes with a pure PDMS base layer to preserve flexibility.}
                \label{fig:CBPDMS_Sensors}
\end{figure}

Carbon black-PDMS closed-cell foam pressure sensor arrays were fabricated using alternating segments of piezoresistive CBPDMS foam and silver-carbon black-PDMS (Ag-CBPDMS) electrodes, as seen in Figure \ref{fig:CBPDMS_Sensors}. A four point probe arrangement was used to create a 4x1 linear array of sensor channels, with each channel defined as the voltage difference between the electrodes on either side of a closed-cell CBPDMS foam block. The overall footprint of the array was 80 mm by 20mm, with each piezoresistive foam sensor channel having a width of 5 mm, and Ag-CBPMS electrodes having a width of 4 mm as seen in Figure \ref{fig:dimensions_top}. Each sensor array comprised three layers, a pure PDMS base layer, a middle active layer that contained the electrodes and CBPDMS foam sensor channels, and a thin CBPDMS foam top layer, as seen in Figures \ref{fig:sensor_top} and \ref{fig:dimensions_front}. The height of the middle active layer was varied to create a thick sensor with a 4.76 mm active layer and overall thickness of 7.80 mm, and a thin sensor with a 3.18 mm active layer and overall sensor thickness of 6.20 mm. To supply current to the sensor array and allow for voltage measurements, ribbon cable ends were embedded in the Ag-CBDPMS electrodes, as seen in Figure \ref{fig:sensor_top}.

\subsection{Sensor Array Fabrication}

\subsubsection{Electrode Layout and Material Selection}
\label{Sec:4point}

\begin{figure}
        \centering
                \includegraphics[width=7cm]{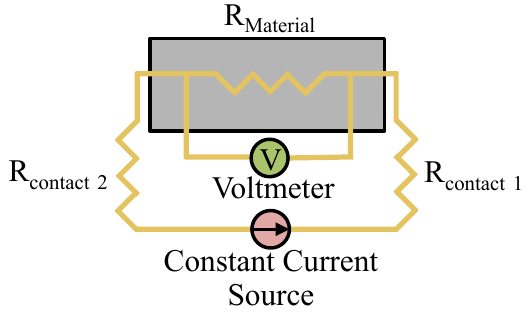}
                \caption[CBPDMS foam array four point measurement]{A four point probe measurement, described in section \ref{Sec:4point}, was used to eliminate the impact of contact resistance on voltage measurements.}
                \label{fig:4point}
\end{figure}

Electrode layout and material selection were driven by the need to eliminate the effects of contact resistance and maximize the conduction between the electrodes and the CBPDMS foam active sensing material while retaining overall array flexibility. To eliminate contact resistance, a four point probe arrangement was utilized. In a four point probe, a pair of outside leads are driven by a constant current source, while inner lead pairs are used to take voltage measurements, as seen in Figure \ref{fig:4point}. In this arrangement, virtually no current flows in the measurement leads, meaning the voltage drop due to the contact resistance between the measurement leads and the CBPDMS foam is negligible, allowing the resistance variation with strain of the CBDMS foam to be isolated. 

\begin{figure}
        \centering
        \includegraphics[width=8cm]{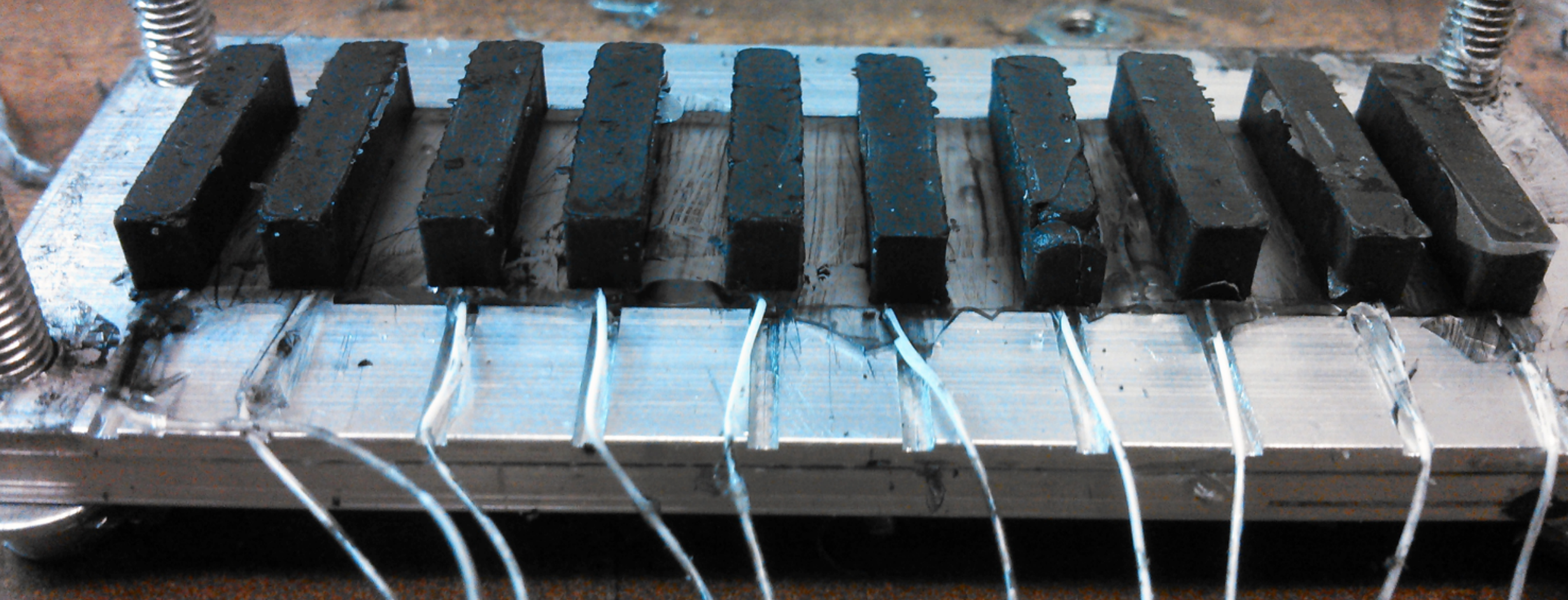}
                \caption[Ag-CBPDMS electrodes]{To maintain overall sensor array flexibility while providing maximum adhesion with the CBPDMS foam piezoresistive material, electrodes were fabricated from a Ag-CBPDMS composite with high conductivity and low piezoresistivity.}
                \label{fig:electrodes}
\end{figure}

When making electrical connections to CBPDMS foam, the contact area between the electrode and the foam sensing material played a large role in the quality of voltage measurements \cite{Dusek:2014p4920}.  To maximize contact between the electrodes and the CBPDMS foam, electrodes spanned the full height of the sensor array and were embedded within the foam, as seen in Figure \ref{fig:electrodes}. To retain array flexibility and maximize adhesion between the electrodes and the CBPDMS foam, electrodes were fabricated from a silver-carbon black-PDMS composite (Ag-CBPDMS). The use of silver and a high carbon black concentration within the electrode ensured a composite with high conductivity that was insensitive to strain. By retaining PDMS as the bulk material in the composite, the electrodes remained flexible and excellent adhesion was observed between the electrodes and CBPDMS foam. 

To allow for communication with the sensor arrays, ribbon cables were embedded in the AG-CBPDMS electrodes. Prior the embedding the stripped wire ends in the composite, the wires were tightly wound into coils, which both maximized the contact area between the wire and the AG-CBPDMS and enhanced the mechanical bond between the components. 

\subsubsection{Electrode Fabrication}
\label{Sec:electrode}

To prepare the AG-CBPDMS, silver powder (Number 11402 from Alfa Aesar) was mixed with PDMS (Sylgard 184 from Dow Corning) and carbon black (from Denka Elastomer and Performance Plastics) at a composition of 72.94\% AG, 3.79\% CB, and 23.27\% PDMS by mass in a Kurabo Mazerustar KK series mixer. The uncured AG-CBPDMS mixture was manually spread into a delrin mold with an aluminum base plate. Wires with pig tails were embedded in each uncured electrode, and the mold was capped with an aluminum top plate before being cured in an oven at 120$^{\circ}$c for approximately two hours. 

Following curing and cooling, excess AG-CBPMS was carefully trimmed using a razor, and the electrodes were individually removed from the delrin mold, allowing any excess material to be trimmed using a razor blade. After the delrin mold was cleaned, electrodes were reinserted, and a 2 mm layer of pure PDMS was applied over the electrodes to serve as a structural base for the sensor array. After curing the pure PDMS in the oven for approximately 20 minutes, the completed electrode layer was able to be removed from the mold, as seen in Figure \ref{fig:electrodes}.

\subsubsection{CBPDMS Foam Fabrication}
\label{Sec:foam}

The CBPDMS foam used in the pressure sensor arrays was fabricated using Soma Foama 15 from Smooth-On. Soma Foama is a two-component platinum silicone casting foam that is available in both 240 kg/m$^3$ (15 lb/ft$^3$) and 400 kg/m$^3$ (25 lb/ft$^3$) densities. The Soma Foama 15 was found to expand 3-4 times its initial volume when mixed, and because the foaming reaction was rapid, advanced preparation of all molds was required to ensure successful fabrication. Soma Foama 15 used a 2:1 ratio of the A:B silicone components by volume or weight, and carbon black particles were mechanically mixed into Part A in advance of the foaming reaction using the Mazerustar mixer at a ratio of 5\% by mass carbon black.

When ready to mold the CBPDMS foam sensing layer, Soma Foama Part B was mixed into the Soma Foama Part A-carbon black mixture by hand, taking care to ensure thorough mixing of the two components. The mixture was then spread into the gaps between the AG-CBPDMS electrodes, and the molds were capped with an aluminum plate. A small gap was maintained between the electrode layer and the top plate, allowing the foam mixture to expand out the sides of mold after the electrode gaps were filled and leaving a uniform layer of foam across the top of the array, as seen in Figure \ref{fig:sensor_top}. After allowing the completed array to cure for approximately 12 hours, the excess foam was trimmed away from the array, and the edges of array were coated in a thin layer of pure PMDS to insulate the exposed electrode edges from the environment, producing the final sensor array shown in Figure \ref{fig:sensor_top}.

\section{CBPDMS Foam Sensor Array Characterization}
\label{Sec:characterization}

To characterize the performance of the CBPDMS foam sensor arrays, a time-varying hydrostatic pressure was created by oscillating the sensor arrays vertically in the water column using a computer controlled linear stage. From these experiments the basic operating characteristics of the sensor arrays including the repeatability and dynamic range were studied, and a sensor calibration was developed. 

\subsection{Sensor Power and Amplification Electronics}
\label{Sec:electronics}

For the CBPDMS foam sensor array characterization experiments, custom electronics were designed to provide a current source for the four-point probe arrangement discussed in Section \ref{Sec:4point}, and to amplify voltage measurements from the array. The current source and amplification circuits were built on a custom PCB board designed and printed for the experiments. Signal amplification was provided for each sensor channel by an AD620 operational amplifier with an amplification gain of 2.66. Following amplification, signals were filtered using an RC low pass filter with a cutoff frequency of 112.88 Hz. 

\paragraph{Constant Voltage Source}
\label{sec:voltage_source}
 
During the plunging experiments, the sensor arrays were inadvertently powered with a constant voltage of 24 volts. The CBPDMS foam sensor array was found to have a total resistance of $\sim$28 k$\Omega$ which can be largely attributed to the contact resistance between the ribbon cables and the Ag-CBPDMS electrodes. With the $\sim$ 28 k$\Omega$ array resistance in series with a 112 k$\Omega$ resistor, the current through the sensor array was $\sim$ 0.17 mA. Because the voltage source was constant in this arrangement, variations in the sensor array resistance could lead to variations in the current being supplied to the array. It was observed that during a typical experiment, the variation in CBPDMS foam resistance could cause an $\sim$ 1\% change in supply current. Because the potential change in current was found to be minor, it will be noted as a source of experimental measurement error, as discussed in Section \ref{sec:verror}. 

\subsection{Plunging Experimental Setup}

\begin{figure}
\centering
\includegraphics[width=8cm]{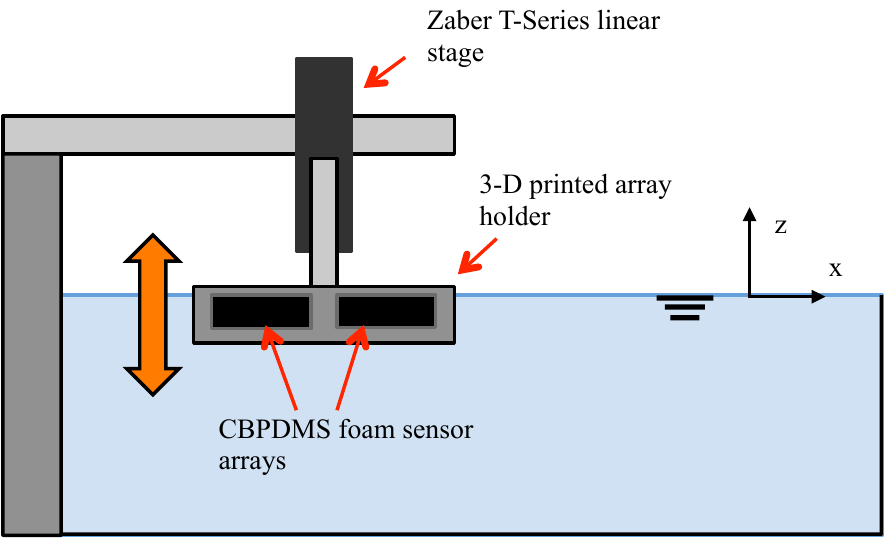}
\caption[Plunging test setup]{Vertical oscillation of the sensor arrays was provided by a computer-controlled Zaber T-Series linear stage mounted above the large vehicle testing tank on the CREATE campus in Singapore.}
\label{fig:zaber_stage}
\end{figure}

To allow for vertical plunging of the sensor arrays, a Zaber T-Series linear stage (Model T-LSQ300B) was mounted vertically above the testing tank, as seen in Figure \ref{fig:zaber_stage}. The CBPDMS foam pressure sensor arrays were mounted in a 3-D printed holder which was attached to the linear stage using an aluminum plate. Because the aluminum plate was thin, oscillation of the sensor holder due to actuator vibrations or fluid interactions was a consideration, but were found to be negligible for the low frequencies (0.5-1 Hz) being investigated.

The motion of the linear stage was controlled using proprietary Zaber Console software running on a laboratory laptop, and connected to the stage via a USB-serial converter. Motion of the Zaber stage was provided by a stepper motor, and stage position and velocity values were reported by the Zaber Console software. The position of the linear stage was written to an output file throughout each experiment to allow the hydrostatic pressure acting on the arrays to be determined. The output voltage from the the pressure sensor arrays was recorded using an NI USB-6218 data acquisition board and National Instruments Labview software running on a laboratory laptop. 

\subsection{Experiment Description and Parameters}

To characterize the dynamic response of the CBPDMS foam pressure sensor arrays, oscillatory plunging motions were used to generate a time-varying hydrostatic stimulus. Because of the limitations of the Zaber stage, and to minimize the impact of spurious vibrations on the experimental results, experiments were restricted to frequencies of 0.5 Hz and 1 Hz. In each experiment, the stage was positioned with the still waterline level with the top of the array, as seen in Figure \ref{fig:zaber_stage}. To allow time synchronization between the stage position and the sensor voltage output, a step in the positive $z$ direction was executed at the beginning of each experiment to create an easily identifiable feature in both signals. Following the step move, ten oscillatory cycles were executed with peak-to-peak amplitudes ranging from 10mm to 30mm. 

The total time-varying pressure ($P_{T}$) during the plunging experiments comprised both a hydrostatic pressure component ($P_{H}$) and a dynamic pressure component ($P_{D}$), as shown in Equation \ref{eq:plunging_pressure}. In Equation \ref{eq:plunging_pressure}, $\phi$ is the velocity potential, $\rho$ is the fluid density, and z is the vertical position of the sensor array midpoint. When considering the time-varying pressure from a simulated sinusoidal motion with a 30mm peak-to-peak amplitude at 1 Hz, the dynamic pressure component was found to be at least two orders of magnitude smaller than the hydrostatic component, and therefore the dynamic pressure component was neglected from future analysis. 

\begin{align} \label{eq:plunging_pressure}
P_{T}&=P_{H}+P_{D} \\
\nonumber P_{H}&=- \rho * g * z \\
\nonumber P_{D}&=-\rho \left( \frac{\partial \phi}{\partial t} + \frac{1}{2} \left|\nabla \phi \right|^2 \right)
\end{align} 

It was observed during the course of the experimental program that the sinusoidal motion created by the Zaber stage was not always consistent in amplitude and frequency. This was particularly evident during low frequency motions were the stage would occasionally hesitate at the peak or trough of a cycle. These slight errors in amplitude and frequency were accounted for by utilizing the actual position output recorded from the stage during each experiment instead of relying on the anticipated frequency and amplitude values. 

\subsection{Plunging Experiment Results}

The voltage output signals from the CBPDMS foam sensor were bandpass filtered using the \emph{idealfilter} function in Matlab. The filter is non-causal with sharp cutoffs in the frequency domain and non-realizable in real time. The bandpass interval was chosen to be 0.2 to 3.0 Hz for the plunging experiments in order to remove any low frequency DC drift from the signals, as well as high frequency electrical noise. 

\subsubsection{Inverse Pressure-Resistance Relationship}
\label{sec:P_R}

\begin{figure}
        \centering
        \subfigure[Plotting the hydrostatic pressure stimulus (black line) and the CBPDMS foam sensor voltage output (solid colored lines) together shows that an inverse relationship exists between hydrostatic pressure and output voltage for the CBPDMS foam sensing material. Because output voltage is proportional to resistance, it follows that the resistance change in the material is inversely proportional to pressure.]{
                \includegraphics[width=8cm]{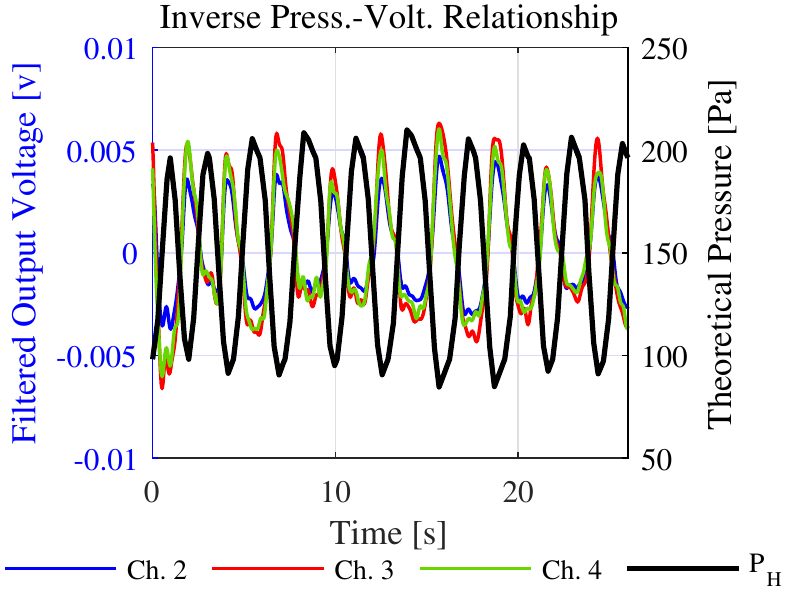}
                \label{fig:sinusoid_plunging_pvsv}
                }
        \subfigure[To form pressure vs. voltage pairs for use in sensor array calibration, the channel voltages were resampled at times corresponding to Zaber stage position outputs (designated by circles, squares, and diamonds). The mean voltage has been removed from each channel voltage.]{
                \includegraphics[width=8cm]{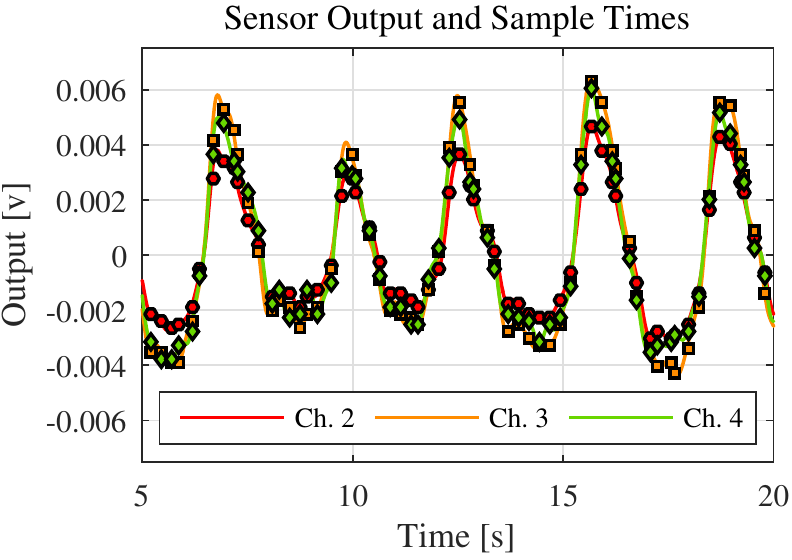}
                \label{fig:v_stagetimes}
                }
\caption[Inverse relationship between resistance and pressure]{The voltage output from the CBPDMS foam sensor array during oscillatory plunging revealed an inverse relationship between pressure and resistance. The voltage time series was sampled at discrete times corresponding with position outputs from the Zaber stage in order to generate pressure-voltage pairs for use in sensor calibration. Sensor array channel 1 was not operational and is not shown.}
\label{fig:zaber_timedomain}
\end{figure}

Figure \ref{fig:zaber_timedomain} shows an inverse relationship exists between voltage output and pressure magnitude when the CBPDMS foam arrays were subjected to an oscillatory pressure stimulus. For a constant current, channel voltage is proportional to resistance through Ohm's law, so it follows that the resistance change in the CBPDMS foam sensing material was inversely related to pressure. The inverse relationship between pressure and resistance follows the compressible model of CBPDMS piezoresistivity discussed in Section \ref{Sec:models}. Throughout extensive testing of the CBPDMS foam sensor array using multiple hydrodynamic stimuli, the inverse relationship between the resistance and pressure remained consistent and repeatable, and it was concluded that for isotropic hydrodynamic stimulus, the CBPDMS foam sensor arrays follow the compressible model of piezoresistivity.

\subsubsection{Sensor Array Repeatability}
\label{sec:repeatability}

To evaluate the repeatability of the CBPDMS foam arrays, voltage was plotted versus hydrostatic pressure for the periodic plunging experiments. To be considered repeatable, the pressure-voltage pairs from multiple plunging cycles should fall on the same curve. 

As mentioned previously, the oscillatory motion of the Zaber stage varied slightly in amplitude and period throughout an experiment. To account for these variations, sensor output voltage values were selected for the times corresponding to the position outputs from the Zaber stage, as seen in Figure \ref{fig:v_stagetimes}. In this way, an accurate picture of the  pressure-voltage relationship for the array could be found.

\begin{figure}
        \centering
        \subfigure[Foam array channel 2 pressure vs. voltage pairs from plunging experiments without DC offset voltage adjustment.]{
                \includegraphics[width=8cm]{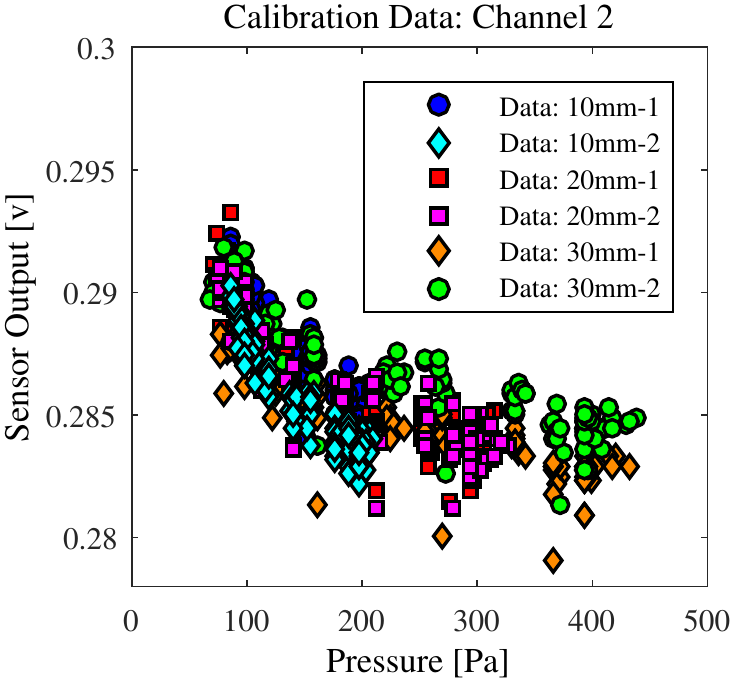}
                \label{fig:raw_PvsV}
                }
         \subfigure[When the DC offset was adjusted for data sets `10mm-2' (light blue diamonds) and `30mm-2' (green circles), the spread in data points between the six plunging experiments was reduced.]{
                \includegraphics[width=8cm]{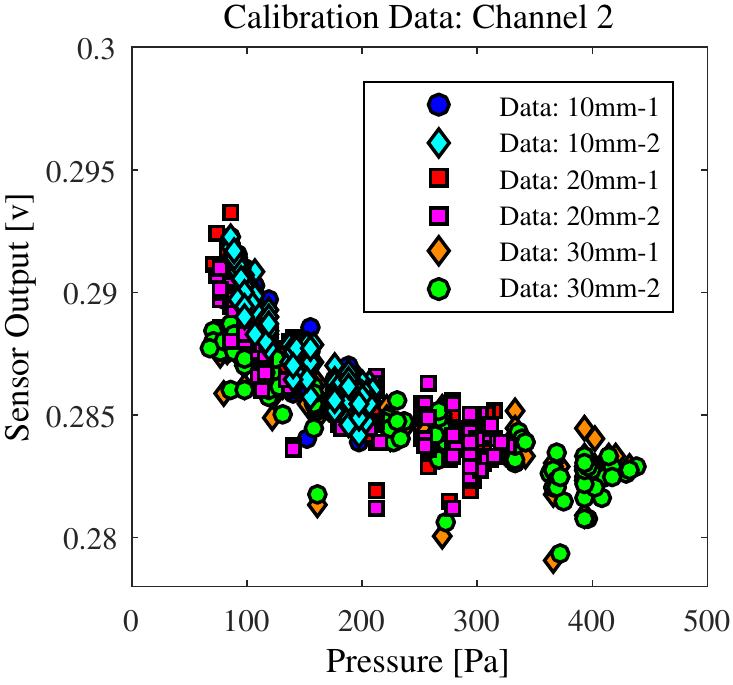}
                \label{fig:shift_PvsV}
                }
                \caption[Pressure vs. voltage plots for 7.8 mm thick foam array in plunging]{Pressure-voltage pairs from independent plunging experiments at 10 mm, 20 mm, and 30 mm peak-to-peak amplitudes (A$_{pp}$). Two experiments were conducted at each amplitude, and are designated as A$_{pp}$-1 and A$_{pp}$-2.}
\label{fig:Thick_PvsV}
\end{figure}

A qualitative assessment of sensor array repeatability was found by plotting the results from six independent plunging experiments with the 7.8 mm thick sensor array on the same pressure vs. voltage axes, as seen in Figure \ref{fig:raw_PvsV}. Within each experiment, the data points were found to follow a consistent trend, qualitatively fulfilling the requirement of repeatability within a given experiment. Although the dynamic response of the sensor array appeared consistent between independent experiments, variations in the DC offset voltage were observed, leading to a distinct spread in the data points.

\paragraph{DC Offset Voltage}

When the results from six independent plunging experiments with the 7.8 mm thick sensor array were plotted on the same pressure vs. voltage axes, it was observed that the DC voltage offset varied between independent experiments, as seen in Figures \ref{fig:raw_PvsV}. Because the CBDPSM foam sensor array is intended for the measurement of AC pressure signals, the dynamic response of the array is of primary concern. In order to verify the repeatability of the dynamic response of the sensor array between independent experiments, the DC voltage offsets were adjusted for consistency, as seen in Figures \ref{fig:shift_PvsV}. With the DC voltage offset corrected, the results from all six independent experiments were found to have a repeatable non-linear pressure vs. voltage response.

\subsubsection{Experimental Error in Plunging Results}
\label{sec:verror}

\paragraph{Error in Voltage Measurement}

\begin{figure}
        \centering
                \includegraphics[width=8cm]{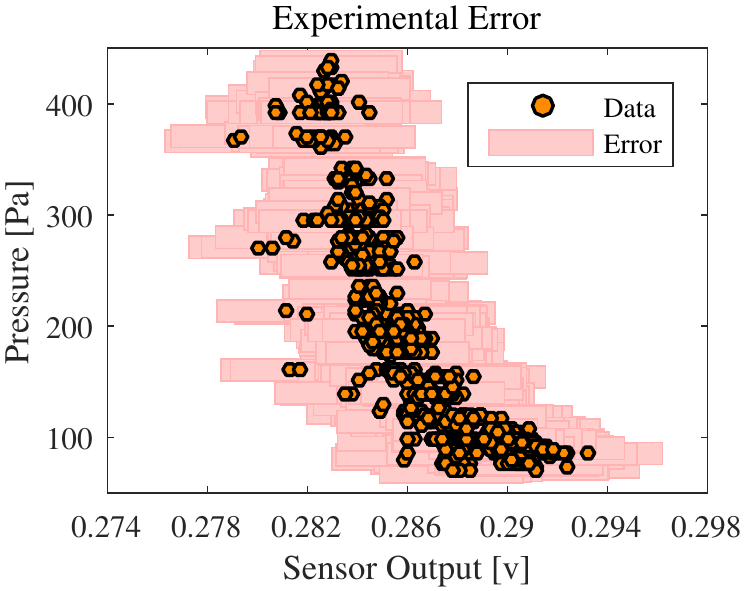}
\caption[Plunging data for 7.8 mm thick array channel 2 with error]{Plunging data for 7.8 mm thick array channel 2. The light red shaded boxes designate the experimental error in pressure and voltage. Data is plotted as voltage vs. pressure to reflect the orientation of the data for calibration of the CBPDMS foam array.}
\label{fig:thick_error_ch2}
\end{figure}

During the plunging experiments, the CBPDMS foam sensor array was powered using a constant voltage source, as described in Section \ref{sec:voltage_source}. It was found that the supply current to the sensor array could vary by $\sim$1\% if the total resistance of the sensor array was a summation of the resistance of each individual electrode and sensor channel resistance. To account for this possible variation, the error in voltage, given by the red shaded boxes in Figure \ref{fig:thick_error_ch2}, was found by scaling each voltage measurement based on a supply current increase and decrease of 1\%, as given in Equation \ref{eq:v_error}. For the 6.2 mm thick (Thin) sensor array, the magnitude of voltage fluctuation was found to be approximately twice that of the 7.8 mm (Thick) array for the same pressure stimulus. 

\begin{equation}\label{eq:v_error}
v_{error}=(1\pm0.01)*v
\end{equation}

\paragraph{Error in Theoretical Hydrostatic Pressure}

The pressure used to generated pressure-voltage pairs in Figures \ref{fig:zaber_timedomain} and \ref{fig:Thick_PvsV} was the hydrostatic component of pressure based on the position output from the Zaber linear stage. Two likely sources of error in the theoretical hydrostatic pressure were the presence of surface waves in the tank generated by the motion of the sensor holder beneath the surface, and measurement error in the initial position of the sensor. 

The possible error in hydrostatic pressure due to free surface fluctuations in the tank was characterized by considering the voltage measurements from the CBPDMS foam array after the motion of the stage had completed it's motion. Because the stage motion was complete and the array was at rest, fluctuations in the voltage measurements were likely due to residual free surface motion in the tank. After applying the calibration described in Section \ref{sec:piecewise} to the raw voltage measurements, the pressure fluctuations were found to be $<\sim$10 Pa for the 30 mm peak-to-peak case, where maximum free surface disturbances were expected. Because the maximum submergence depth in the plunging experiments was 40 mm, the decay of the pressure fluctuations due to free surface deformation was found to be small, and a constant $\pm$10 Pa error was applied to the pressure values in Figure \ref{fig:thick_error_ch2}. Additional error in the theoretical hydrostatic pressure may have been present due to errors in the initial measurement of sensor array position, and measurement error was assumed to be of similar magnitude or less than the $\pm$10 Pa error assumed for free surface deformations.

\subsection{Calibration of 7.8 mm (Thick) CBPDMS Foam Array}

\subsubsection{Polynomial Fits for Full Data Set}
\label{sec:fits_full}

Using the curve fitting function \emph{fit} in Matlab, robust polynomial fits of order 2-4 were found for the 7.8 mm thick CBPDMS foam plunging data, as seen in Figure \ref{fig:poly234_woutliers} for sensor channel 2. It was observed that while the 2nd (red) and 3rd (black) degree fits were very similar in shape, the 4th degree fit displayed non-physical characteristics at the high-pressure extreme of the calibration data range due to the influence of outliers at particularly low voltage values. The RMS error between the expected pressure values and polynomial fit, RMS error normalized by the pressure range (Equation \ref{eq:nrmse}), and $R^{2}$ values can be found in Table \ref{tab:fits_woutliers_ch2}. Based on the results for second, third, and fourth degree polynomials, it was decided that the third degree polynomial provided the best representation of the voltage-pressure relationship without the end effects observed for the fourth degree fit. 

%\begin{equation}\label{eq:rmse}
%RMSE=\sqrt{\frac{1}{n}\sum_{i=1}^{n}w_i \left( \hat{y_i} - y_{i}\right)^2} 
%\end{equation}

\begin{equation}\label{eq:nrmse}
NRMSE=\frac{RMSE}{P_{max}-P_{min}} \times 100 
\end{equation}

%\begin{equation}\label{eq:r2}
%R^{2}=\frac{\sum_{i=1}^{n}w_i \left( \hat{y_i} - \bar{y}\right)^2}{ \sum_{i=1}^{n}w_i \left({y_i} - \bar{y}\right)^2}
%\end{equation}

\begin{table}
    \begin{center}
        \begin{tabular}{|lccccc|}\hline
        n & $R^{2}$ & RMSE [Pa] & NRMSE [\%] & &\\
        \hline
        2 & 0.8823 & 33.73 & 9.14 & & \\
        3 & 0.8853 & 33.32 & 9.02 & & \\
       4  & 0.8882 & 32.92 & 8.92 & & \\
       \hline
        \end{tabular}
        \caption[Thick channel 2 calibration]{Goodness of fit measures for the 7.8mm CBPDMS foam sensor array Channel 2 with outliers included.}\label{tab:fits_woutliers_ch2}
    \end{center}
\end{table}

\subsubsection{Identification and Removal of Outliers}
\label{sec:outliers}

When considering the CBPDMS foam array plunging data, outliers were considered samples that fell greater than 1.5 standard deviations away from the third degree polynomial fit found using the complete data set in Section \ref{sec:fits_full}. In Figure \ref{fig:thick_outliers}, the points that satisfied the outlier criteria are designated with large red asterisks, and were removed from the training data set when generating the refined curve fits shown in black. The third degree polynomial coefficients and the goodness of fit metrics for the refined curves can be found in Table \ref{tab:coeffs_refined}. The outliers present in the experimental data were attributed to the first oscillation of each plunging experiment which consistently produced a lower sensor output voltage than subsequent oscillations. The source of this transient behavior in the output voltage is unknown, an considered as a direction for future study.

\begin{table}
    \begin{center}
        \begin{tabular}{|lcccc|}\hline
        Channel & $R^{2}$ & RMSE [Pa] & NRMSE [\%] &\\
        \hline
        2 &   0.8881 & 32.80 & 8.89 &  \\
        3 &  0.8551 & 37.54 & 10.17 & \\
        4  &0.8752 & 34.49 & 9.34 &  \\
        \hline   
        \end{tabular}
        \caption[Refined 3rd order calibration for thick array]{Refined third degree polynomial goodness of fit measures for the 7.8mm CBPMDS foam sensor array}\label{tab:coeffs_refined}
    \end{center}
\end{table}

\subsubsection{Piecewise Calibration with First-Order Correction} 
\label{sec:piecewise}

\begin{figure}
        \centering
        \subfigure[Second, third, and fourth order polynomial curve fits were found for CBPDMS foam sensor array channel 2.]{
                \includegraphics[width=7cm]{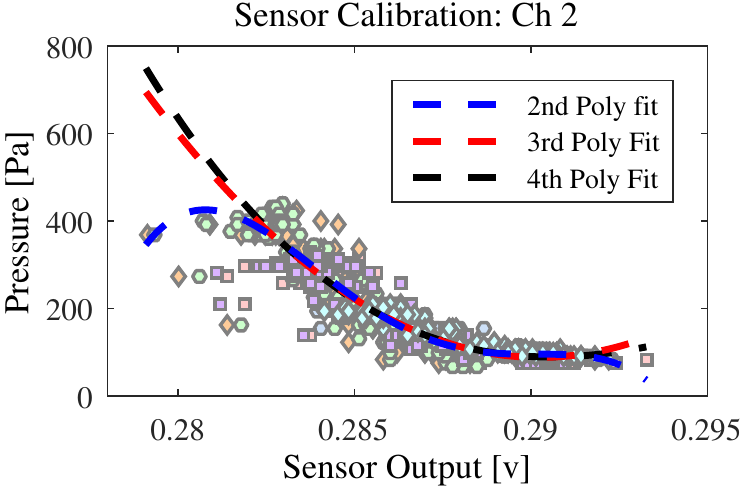}
                \label{fig:poly234_woutliers}
                }
         \subfigure[Outliers were designated as samples falling further than 1.5 standard deviations away from the third degree polynomial fit.]{
                \includegraphics[width=7cm]{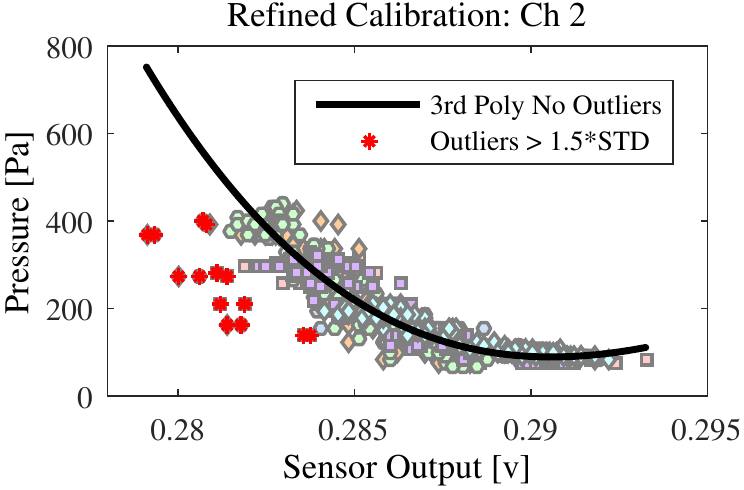}
                \label{fig:thick_outliers}
                }
                \subfigure[Piecewise polynomial calibration for 7.8 mm thick CBPDMS foam array Channel 2.]{
                \includegraphics[width=7cm]{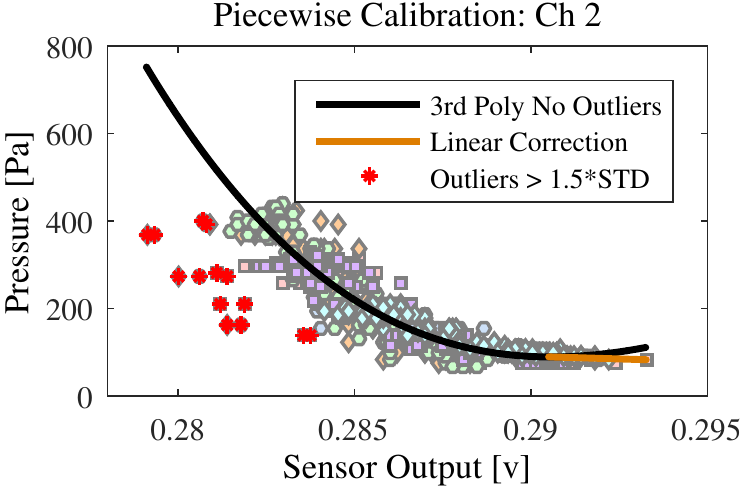}
                \label{fig:piecewise_fits}
                }
\caption[Identification of outliers]{The CBPDMS foam sensor arrays were calibrated by refining a third degree polynomial fit through the elimination of outlier points and the addition of a linear correction for the low pressure regime.}
\label{fig:thick_calibration}
\end{figure}

The third order polynomial curve shown for channel 2 in Figure \ref{fig:thick_outliers} was observed to exhibit non-physical behavior at the low pressure and high voltage end of the training data set. While the slope of the polynomial curve changes from negative to positive, the pressure-voltage relationship in the physical sensor is always inverse with a sensitivity that decreases at high levels of strain (pressure). To better represent the physical performance of the sensor array, a first-order correction of the form shown in Equation \ref{eq:cal_piecewise} was added to the calibration curve in the low pressure region to create a piecewise calibration.

\begin{align}\label{eq:cal_piecewise}
P&=C_{3}*v^{3}+C_{2}*v^{2}+C_{1}*v+C_{0}, v \leq v_{in}\\
\nonumber P&=D_{1}*v + D_{0}, v > v_{in}
\end{align}

To determine the curve fit for the first-order correction, the inflection point ($v_{in},P_{in}$) in the refined third degree polynomial fit was found. A linear fit was then performed on the subset of the data for $v > v_{in}$ with a condition put on the linear fit that it needed to pass through the point ($v_{in},P_{in}$) from the third degree polynomial. The linear fit was found using the \emph{slmengine} function in Matlab which allows for curve fitting with prescriptions on shape, piecewise segments, etc. The calibration curve with first and third degree polynomial regions for channel 2 of the 7.8 mm thick CBPDMS foam array can be seen in Figure \ref{fig:piecewise_fits}, and the calibration curve parameters are given in Table \ref{tab:piecewise_param}.

\begin{table}
    \begin{center}
        \begin{tabular}{|lcccc|}\hline
        Ch. & $C_{3}*10^{7}$ & $C_{2}*10^{7}$  &$C_{1}*10^{7}$ & $C_{0}*10^{6}$ \\
        \hline
        2 & -14.3& 12.8 & -3.81& 3.79 \\
        3 & 1.15 & -1.24 & 0.43 & -0.46 \\
        4 & -28.5 &  32.0  & -12.0 & 15.0 \\
         \hline 
         Ch. & $v_{in}$ [v]  & $P_{in}$ [Pa] & $D_{1}$ & $D_{0}$ \\
         \hline
         2 & 0.291 & 89.25 & -2362.43 & 775.74 \\
         3 & 0.432 & 86.46 & -553.09 & 325.23 \\
         4 & 0.376 & 84.99 & - 3437.60 & 1377.53 \\
        \hline
        \end{tabular}
        \caption[Piecewise polynomial calibration for thick array]{Coefficients and break points for the piecewise polynomial calibration of the thick array. For channel 4, an inflection point did not occur within the training data range, so a linear correction was added for v $>$ 0.376 to ensure extrapolated values followed expected behavior.} \label{tab:piecewise_param}
    \end{center}
\end{table}

\begin{table}
    \begin{center}
        \begin{tabular}{ |l|c|c|c|c|}
        \hline
        Ch. & Exp. & RMSE [Pa]  & NRMSE [\%] \\
        \hline 
        2 & 10 mm & 19.78 & 12.84 \\
        2 & 20 mm & 35.20  & 12.73 \\
        2 & 30 mm & 47.41  & 13.34 \\
        \hline
        3 & 10 mm & 22.04 & 12.52 \\
        3 & 20 mm & 47.77 & 15.09 \\
        3 & 30 mm & 55.08  & 16.43 \\
        \hline
        4 & 10 mm & 32.58  & 15.27\\
        4 & 20 mm & 38.14  & 12.66 \\
        4 & 30 mm &  71.82  & 24.92 \\
        \hline
        \end{tabular}
        \caption[Calibrated plunging results using refined piecewise polynomial calibration for the thick array]{RMS error and normalized RMS error for the piecewise polynomial calibration applied to time series plunging results for the 7.8 mm thick array.}\label{tab:piecewise_results}
    \end{center}
\end{table}

\subsection{Application of Calibration Curve to Time Series Data}
\begin{figure}
        \centering
                \includegraphics[width=2.75in]{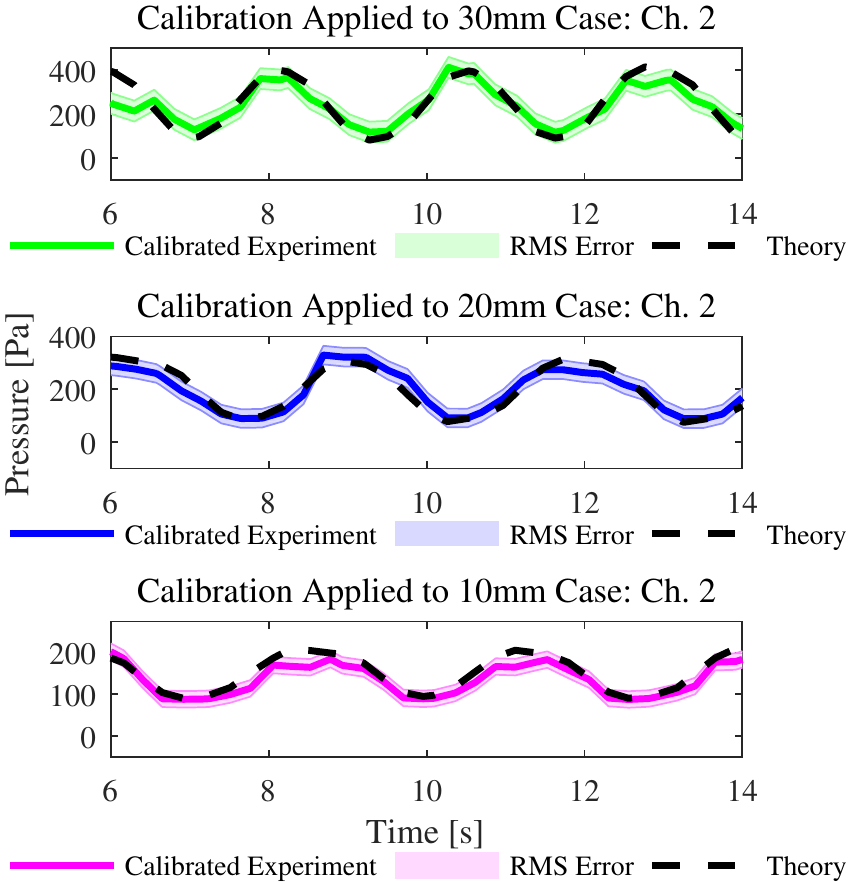}
                \label{fig:calib_timeseries_ch2}
\caption[Polynomial calibration applied to plunging data]{The piecewise polynomial calibration was applied to experimental time series data from plunging experiments (solid colored lines) and compared to the theoretical hydrostatic pressure based on the stage position (black dashed lines) for sensor array channel 2. }
\label{fig:plunging_results}
\end{figure}

The piecewise polynomial calibration curves found for each of the three operational sensor array channels were tested by applying the calibration to the voltage output measurements from plunging experiments at 10 mm, 20 mm, and 30 mm peak-to-peak amplitude, as seen in Figure \ref{fig:plunging_results}. The RMS error (RMSE) and Normalized RMS error (NRMSE) were then calculated between the theoretical hydrostatic pressure determined by the vertical location of the Zaber stage, and the calibrated sensor results, as given in Table \ref{tab:piecewise_results}. The first oscillation of the stage was removed from each data set as discussed in Section \ref{sec:outliers}. Using the piecewise polynomial calibration curves, time-varying hydrostatic pressure was reproduced from sensor voltage measurements with $\sim$15\% NMRSE for the 7.8 mm thick sensor array, and $\sim$13\% NMRSE for the 6.2 mm thick array when compared to the theoretical hydrostatic pressure based on sensor depth.

\section{Conclusions}

Conformal pressure sensor arrays for use in a marine environment were fabricated using a silicone-based material set. The pressure sensitive components of the array utilized a closed-cell piezoresistive CBPDMS foam, while the sensor array electrodes were composed of a silver-carbon black-PDMS composite that displayed high conductivity with low piezoresistivity. The array utilized a pure PDMS base layer to provide support during the fabrication process, and to allow for surface mounting using silicone adhesives. The array utilized a four-point probe measurement technique to mitigate the effects of contact resistance, and electrical connections were provided by a ribbon cable embedded in the Ag-CBPDMS electrodes. Overall sensor array dimensions were 80mm x 20mm with thicknesses of 7.8 and 6.2 mm, and four sensor channels were contained in the array.

Sensor characterization experiments using time-varying pressure from vertical plunging revealed fundamental characteristics of the CBPDMS foam array performance. Primary among these was the non-linear and inverse relationship between hydrodynamic pressure and sensor voltage output that is consistent with the compressible model of CBPDMS piezoresistivity presented in Section \ref{Sec:models}. The response of the sensor arrays across multiple plunging experiments with increasing amplitude was found to be repeatable, and using the results from the plunging experiments, a piecewise polynomial calibration was developed for each channel of the 7.8 mm and 6.2 mm thick arrays.

The success of the CBDPMS foam arrays as a hydrodynamic sensor for use on unmanned marine vehicles can be evaluated by comparing the performance of the arrays during plunging to results from previous studies using distributed arrays of commercially available sensors. Based on previous experiments in vortex tracking on an AUV shaped body \cite{Fernandez:2011p4634} and a single-element hydrofoil \cite{Dusek:2011p4406}, sensor spacing of less than 5 cm was recommended in order to properly resolve vortex location. The spacing of individual sensor channels in the CBPDMS foam arrays is 1.7 cm, satisfying the sensor spacing guidelines for the unmanned vehicle applications presented. From previous experiments, example hydrodynamic pressure stimuli were found to have dynamic pressure amplitudes between 10 and 200 Pa. During the plunging experiments, the dynamic range demonstrated by the CBPDMS foam pressure sensor array was approximately 50-500 Pa, with a resolution of $\sim$10 Pa. In these experiments, the low end of the dynamic range was limited by the experimental setups and the minimum depth necessary to fully submerge the sensor array. 

For use in real-world ocean engineering applications, CBPDMS foam sensor arrays must be robust enough for prolonged exposure to fluids and able to withstand repeated handling and potential impacts. The CBPDMS foam sensor arrays discussed in this study were surface mounted on both flat and curved surfaces using double sided tape and silicone adhesive. To specifically test the robustness of the 7.8 mm thick CBPDMS foam array, the array was left submerged in the small towing tank at MIT for over 24 hours, and no impact  was observed on performance.

To bring distributed pressure sensing with high spatial resolution to full-scale ocean engineering systems, pressure sensors must be low cost. The cost of current commercially available sensors was illustrated by the unmanned kayak vehicle which was equipped with an array of 20 Honeywell SPT series sensors at a total cost of $\sim$\$5000, or $\sim$\$250 per sensor \cite{Dusek:2013p4635}. The CBPDMS foam sensor arrays were fabricated using approximately $\sim$\$40 worth of materials for a 4x1 array, or $\sim$\$10 per sensor. With modifications to the sensor electronics and data acquisition technique, the array could be converted to a 7x1 array, reducing the cost per sensor to $\sim$\$6. The largest contributor to the material cost was the silver used in the fabrication of the sensor array electrodes. With a modification to the materials used in the electrodes, the sensor cost could be reduced further, representing a greater benefit over commercially available sensors for large-area distributed applications.

\section{Acknowledgments}
The authors would like to acknowledge the support of a grant from NOAA's MIT SeaGrant program, The Singapore-MIT Alliance for Research and Technology (SMART), and the William I. Koch chair for making this work possible. Special thanks to Matthew D'Asaro and Jessica Herring for their help in the development of doped-silicone sensor materials, and Vignesh Subramaniam for his experimental assistance at the CENSAM laboratory and testing tank in Singapore.

\section*{References}

\bibliography{DusekSA}

\end{document}